\documentclass[twocolumn,aps,floats,superscriptaddress,prl]{revtex4}
\usepackage{epsfig}
\usepackage{amsmath}
\usepackage{graphicx}
\usepackage{bm}

\begin{document}

\title{Effect of weak disorder induced on the first order transition in a p-wave superconductor.}

\author{Priyanka Mohan}
\author{R. Narayanan.}
\affiliation{Department of Physics, Indian Institute of Technology Madras,
Chennai 600036, India}
\date{\today}

\begin{abstract}
 We investigate the influence of weak 
disorder on the fluctuation driven first order transition found in a p-wave superconductors.
The flow equations for the disorder averaged effective theory is derived in an 
$4-\epsilon$ expansion. The fixed point structure of these flow equations and their 
stability is discussed.
\end{abstract}
\maketitle

The role of quenched disorder near second order phase transitions have been the subject 
of intense investigations in the last few decades. These investigations led to the 
development of the so-called Harris criterion \cite{harris} which helps to gauge the relevancy 
of weak disorder  near the clean critical point.  
Also, there is a significant body of work that deals 
with strong disorder effects near various critical points. 
These are generically termed as Griffiths phenomenon \cite{griffiths}, which
leads to very interesting behavior of various observable quantities like the 
susceptibility in the vicinity of critical points. 
However, the role played by quenched impurities near first order transitions has not been so thoroughly investigated. A notable 
exception  to the above statement is the pioneering work of
Aizenmann and Wehr, \cite{aizenmann} where it was proved that in dimensions,  $d\le2$ the
addition of quenched disorder leads to rounding of the transition. Hui and Berker, \cite{hui},  also came up with exactly the 
same conclusion using heuristic Renormalization Group (RG) arguments. 

    The case in higher dimensions is not so clear. In other words, in certain scenarios, the presence of 
    quenched disorder rounds the first order transition and in other cases it seems that there is no 
  impact on the order of the transition. 
  For instance, the first order transition inherent  in a magnetic system with cubic anisotropy 
 can be shown to be unaffected by the addition of weak-disorder \cite{rn_vojta}.
 
 An example in higher dimension wherein disorder does alter the nature of the transition 
 is given by the case of type-II superconductors \cite{halperin}: In the case of a clean
 type-II superconductor Halperin and co-workers had shown that the transition was of the 
 fluctuation driven first order type. This conclusion was based on a $4-\epsilon$ expansion, 
 wherein the RG flows showed runaway behavior. These results for the clean superconductor 
 were also in accordance with the results obtained by Coleman and Wienberg \cite{coleman}
 for the same model. 
 In Ref. \cite{cardy},  Boyanovsky and Cardy analyzed the 
  effect of impurities on the fluctuation driven first order transition in such a s-wave superconductor. 
  In particular it was shown in Ref. \cite{cardy}, that if one adds quenched disorder as a random mass term, 
  then as long as the "bare disorder" strength is above a threshold 
 value the first order transition inherent in scalar electrodynamics, (type-II superconductor), gets converted into a 
 second order transition. In other words, if one starts with a bare disorder value that is below the threshold value 
 then the first order transition persists. 
 These results were obtained in an $\epsilon=4-d$ expansion.  
 
 In similar vein, using the $\epsilon = 4-d$ expansion, Blagoeva, et al., \cite{blag}, 
 studied the nature of the transition in an unconventional 
 superconductor with orthorhombic and cubic anisotropies: They found that the phase transition 
 in these superconductors was predominantly weakly first order. However, in a situation where the 
 orthorhombic anisotropy is absent, they found a new critical fixed point. In \cite{bus}, the influence of 
 weak randomness on the fluctuation driven first order transition in such unconventional superconductors were studied. They concluded that in an $\epsilon = 4-d$ expansion the transition 
 in such anisotropic unconventional superconductors remained weakly first order.
 
  Thus, from the discussions of the preceding paragraphs it can be seen that in $d>2$, the influence of 
 quenched disorder on first order transitions is quite model dependent. 
 
 In this brief communication, we look at the effect of quenched non-magnetic impurities on the 
 critical behavior of a p-wave superconductor: 
 The clean (impurity-free), version of this model was recently studied by 
  Qi Li and collaborators \cite{qili}. In an $\epsilon = 4-d$ expansion, 
  when the number of order parameter components, $m$,  are small,
  they showed that the phase transition in such an unconventional superconductor is also
 of the fluctuation driven first order variety. These results were analogous to the results obtained for the case of the s-wave superconductor 
 \cite{coleman}, \cite{halperin}, that we had elucidated earlier.  
 To incorporate the effect of disorder, we start with a random mass 
 generalization of the effective action derived by Qi Li and collaborators 
for the p-wave superconductor. Under this generalization, we set the distance to 
criticality $t \rightarrow t-\delta(t({\bm x}))$, and we obtain:
\begin{eqnarray}
S &=& \int d{\bm x}\ \Bigl[ \bigl(t-\delta(t({\bm x}))\bigr)\vert{\bm\psi}\vert^2 + c\vert{\bm
D}{\bm\psi}\vert^2 + u\vert{\bm\psi}\vert^4 +\nonumber\\
&&v\vert{\bm\psi}\times{\bm\psi}^*\vert^2 +
 \frac{1}{8\pi\mu}({\bm\nabla}\times{\bm A})^2 \Bigr].
\label{eq:1.1}
\end{eqnarray}
Here,  ${\bm D} = {\bm\nabla} - i 2e{\bm A}$, is the minimal coupling term. ${\bm A}$
 is the vector potential, and 
$\bm\psi$, is the $m$ component complex order parameter field. 
We integrate out the quenched disorder by using the replica trick. The replica trick essentially results 
in re-writing the $F= -{\{\log Z\}}_{\rm dis} = \lim_{n \rightarrow 0} \frac{\{{ Z^n\}_{\rm dis}}-1}{n}$. Here, the subscript just implies that one has to perform a disorder average. Now, for calculational ease we 
assume that the disorder distribution is delta correlated. In other words,  we assume that the 
$\{\delta t(x)\}_{\rm dis} = 0$, and $\{\delta t(x) \delta t(y)\}_{\rm dis} = \Delta \delta(x-y)$. We now perform the disorder average and then perform a $1$-loop RG calculation. The result of which this analysis gives \cite{pr}

  \begin{eqnarray}
\frac{du}{dl} &=& \epsilon u+6e^2 u-2(m+4) u^2+8uv-8 v^2+ 12 u \Delta -3 e^4  \nonumber\\
\frac{dv}{dl} &=& \epsilon v+6e^2 v+2m v^2-12uv+ 12 v \Delta   \nonumber\\
\frac{d \Delta}{dl} &=& \epsilon  \Delta+6e^2  \Delta+8  \Delta^2-4(m+1)u \Delta+ 8 v \Delta \nonumber\\
\frac{d e^2}{dl} &=&  \epsilon e^2  -\frac{m}{3} e^4 
\label{eq:1}
\end{eqnarray}

The flow equations, Eq. \ref{eq:1}, appropriately rescaled, maps onto the case of the disordered s-wave superconductor studied in 
Ref. \cite{cardy} when $v=0$. In the same spirit, it is identical to the flow equations derived in Ref. \cite{qili}, for the p-wave superconductor 
when $\Delta =0$. 
In what follows, we analyze the fixed points of the above set of equations. We will linearize the flows around the fixed points and thus discuss their stability. Throughout the course of this paper we have 
suppressed the flow of the distance to criticality $t$. This is on account of the fact that the $t$ is always a relevant operator.  
Thus, we look at only the flow in the $(u,v,\Delta)$ space.  Furthermore, since we are
 interested in the superconducting transition, we will look for a fixed point with finite value of $e^2$.

 \noindent
 {\bf Case 1 :} $ u^*= v^*=\Delta^*=0$. This is the Gaussian fixed point which has eigenvalues $\lambda_u
 =\lambda_v = \lambda_{\Delta}=4-d$. This implies that the Gaussian fixed point is unstable for all 
 $d<4$, consistent with the naive power-counting.
 
 Next we look at a class of fixed points where one of the coupling constants is 
 constrained to a non-zero fixed point value. 
 Now,
 the three such fixed points are :\\
 
 \noindent
 {\bf Case 2 :} $u^*\neq0$, $v^*=0$, and $\Delta^*=0$. 
  \begin{eqnarray}
  \nonumber
  u^* &=& \frac{\epsilon (m+18)}{4 m (m+4)}\left[1\pm\left(1-\frac{216 (m+4)}{(m+18)^2}\right)^{1/2}\right] \\
  \nonumber 
  v^* &=& 0\\
  \nonumber
  \Delta^* &=& 0
    \end{eqnarray}
 Consistent with Refs. \cite{qili}, \cite{halperin}, we 
 call this pair of fixed points the pure s-wave fixed points. It can be shown that this fixed point is 
 physical, (i.e., $u^*$ is real), only when $m > m_c \approx 183$. 
It can be shown that for physically relevant parameter values, i.e., $m>m_c$, this pair of fixed points are 
unstable. The most obvious route to show the instability of the pure s-wave fixed point is to 
calculate the eigenvalue $\lambda_v$, which can be shown to be:  $\lambda_v = \frac{1}{u^*} (2(m-2){u^*}^2+\frac{27 {\epsilon}^2}{m^2})$.  
As can be seen this eigenvalue is always positive in the physically relevant parameter regime.

\noindent
 {\bf Case 3:}  $u^*=0, v^*\ne0$ and $\Delta^*=0$
  \begin{eqnarray}
  \nonumber
u^*&=&0\\
v^*&=&-\frac{\epsilon}{2m}\left[1+\frac{18}{m}\right]\nonumber\\
\Delta^*&=&0  \nonumber
  \end{eqnarray}
  Once, again this is an unstable fixed point. It can be shown that one of the eigenvalues, namely 
  $\lambda_{\Delta}= -2(m-4){v^*}$, is positive for all $m>4$.  Now, for $m<4$, even though $\lambda_{\Delta}$ switches signs, 
  it can be shown by following Ref. \cite{qili}, that this fixed point is 
  unstable.
 
  \noindent
  {\bf Case 4}: $u^*=0, v^*=0$ and $\Delta^*\ne0$
   \begin{eqnarray}
   u^*&=&0\nonumber\\
v^*&=&0\nonumber\\
\Delta^*&=&-\frac{\epsilon}{8}\left[1+\frac{18}{m}\right]\nonumber
\end{eqnarray}
  This is an un-physical fixed point, as $\Delta^*$, which is related to the width of the disorder 
  distribution, is negative at this fixed point. Since, this is an un-physical fixed point, we will refrain 
  from discussing its stability here. 
    
 Now, we turn our attention to the case where two of the three coupling constants attains non-zero fixed point values:

       \begin{figure}
\begin{center}
 
\end{center}\includegraphics[width=8.0cm]{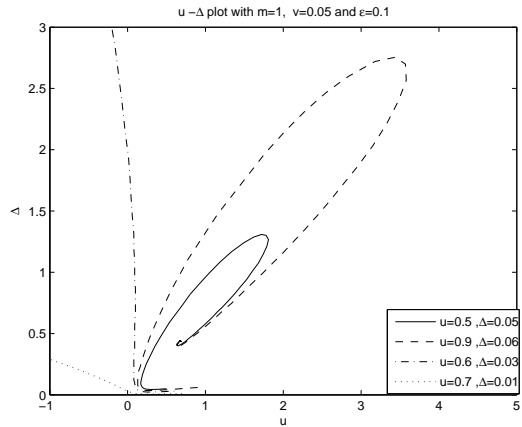}
\caption{Numerical solutions of the RG equations in the $u-\Delta$  plane. The plot is for the case $m=1$, the only case where there is a fixed point for low enough number of order parameter components. It is clearly shown that when 
there is a sepratrix in this set of equations. In other words, when the bare disorder is less than some threshold value then one flows away from the fixed point. However, when the disorder is greater than 
the threshold value, we flow into the fixed point.  These are consistent with the results of \cite{cardy}. }
\label{fig:1}
\end{figure}
 
 \noindent
    {\bf Case 5:} $u^*\ne 0, v^*=0$ and $\Delta^* \ne 0$
  \begin{eqnarray}
   u^*&=&\frac{\epsilon}{8(2m-1)}\left[\left[1+\frac{18}{m}\right]+\sqrt{\left[1+\frac{18}{m}\right]^2+\frac{864(2m-1)}{m^2}}\right]\nonumber\\
     v^*&=&0 \nonumber\\
    \Delta^*&=&\frac{(m+1)}{2}u^*-\frac{\epsilon}{8}\left[1+\frac{18}{m}\right]    \nonumber
  \end{eqnarray}
 Since the fixed point value $v^*=0$, we will call this case the disordered s-wave fixed point.  It is seen from following Ref. \cite{cardy}, that the $\lambda_u$ and 
 $\lambda_{\Delta}$ are both negative for small number of order parameter components. Thus, we need to study the eigenvalue $\lambda_v$ that we 
 obtain from linearizing around the $v^* = 0$ fixed point. It can be shown that $\lambda_v = (4(m-2)u^*+4\Delta^*)$. It is obvious that only for the case 
 $m=1$, will this eigenvalue be negative. Thus, implying that this fixed point is a critical fixed point for $m=1$. For all other values of $m$, this fixed 
 point is unstable and the flow is away from this fixed point. It is of little surprise that at the disordered s-wave fixed point, the operator 
 $v$ is irrelevant, as it can be shown that the $v$ term in the action is identically 
 zero. Thus, for $m=1$, we flow into a critical fixed point first obtained in Ref. \cite{cardy}. As can be seen from Fig. \ref{fig:1}, there is a minimum critical 
 value of the bare disorder strength beyond which the system goes through a second-order transition. Else, the transition remains first order.
 
   
   
 \begin{figure}
\includegraphics[width=8.0cm]{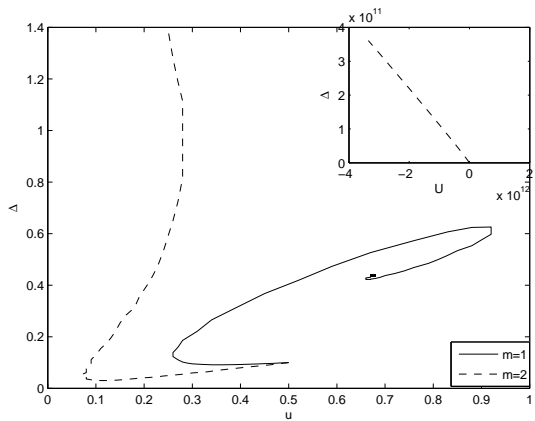}
\caption{ Flow plot from RG equations for two different values of m, $m=1$ and $m=2$ . Both flows start from the same initial value of $u=0.5$, $v=0.04$ and $\Delta =0.1$. For $m=1$, the flow is into the fixed point given by $u^*=0.675$ and $\Delta^*=0.4375$. But for $m=2$, as the graph clearly shows there is runaway flow towards large values of $u$ and $\Delta$. In fact as seen in the inset for $m=2$, the flow crosses over into the classically unstable regime, $u<0$.}
\label{fig:2}  
\end{figure}
 
 \noindent
  {\bf Case 6:}  $u^*\ne 0, v^* \ne 0$ and $\Delta^*=0 $
  \begin{eqnarray}
  u^* &=& \frac{m v^*}{6}+ \frac{\epsilon}{12}\left[ 1+\frac{18}{m}\right]\nonumber \\ 
  v^* &=& -\frac{\epsilon}{2m}\frac{\left[(m-3)(m+4)(m+18)\pm 3M_1\right]}{D_1}\nonumber\\
  \Delta^*&=& 0\nonumber
  \end{eqnarray}
Where, $M_1=(m^4 -204 m^3-1356 m^2-864 m-15552)^{1/2}$ and $D_1= m^3+ 4m^2-24 m+144$.
 It can be shown by following the arguments in \cite{qili},  that such types of fixed points are only 
 physical when $m \ge m_c^{\prime}$. Here, $m_ c^{\prime} \approx 210.45$. 
 
 Furthermore as argued in \cite{qili} we can 
 show that when $m>m_c^{\prime}$, we get a pair of such fixed points  with the one with the higher numerical value of $v^*$ and $u^*$, being the 
 one with negative eigenvalues $\lambda_u$ and $\lambda_v$.  Since, the fixed point 
 with the lower numerical value of $u^*$ and $v^*$, have been shown to be manifestly unstable 
 in \cite{qili}, we will ignore it from our future discussions.
 In analogy with \cite{qili}, the fixed point with larger value of $u$ and $v$ will be called 
 the pure p-wave fixed point.  

 Now, all that remains to be seen is whether this pure p-wave fixed point is 
 stable to infinitesimal perturbation in $\Delta$.  This can be easily accomplished by evaluating $\lambda_{\Delta}$, which has the form $4(2-m)u^*+ 2(4-m)v^*$. By substituting the functional form for 
 the fixed point into the expression for $\lambda_{\Delta}$, we can easily show that that $\lambda_{\Delta}$ is negative for $m>m_c^{\prime}$. These
  expressions are very cumbersome and thus we will 
 discuss our results for particular number of order parameter components. For instance, for the case 
 $m=300$, we can show that the eigenvalues are: $-1.04576\epsilon, -0.537164\epsilon$ and $-0.52356\epsilon $ which clearly indicates that the fixed point is stable.
 Similarily when $m \rightarrow \infty$, it can be shown that the eigenvalues are all negative 
 at the clean p-wave fixed point. 
 Thus, the pure $p$-wave fixed point is stable for large number of order parameter components.
 
\noindent
 {\bf Case 7:}  $u^*=0, v^*\ne 0$ and $\Delta^*\ne 0 $
     \begin{eqnarray}
     u^* &=&0 \nonumber\\
     v^*&=& \frac{\epsilon}{4(m-6)}\left[1+\frac{18}{m}\right]\nonumber\\
     \Delta^* &=& -\frac{\epsilon (m-4)}{8 (m-6)}\left[ 1+\frac{18}{m}\right]\nonumber    
     \end{eqnarray} 
     As can be seen from the above set of equations, the fixed point value $\Delta^*$ is positive only for $m=5$. Thus, it is sufficient to discuss the stability at this 
     particular value of $m$. For $m=5$, it can be shown that the eigenvalues are $29.6994\epsilon$, $-7.94968\epsilon\pm 4.77846\epsilon i$ and $-\epsilon$. Thus, implying that the fixed 
     point is once again unstable. 
   
\noindent
{\bf Case 8:} $u^*\ne0, v^*\ne 0$ and $\Delta^* \ne 0$
\begin{eqnarray}
u^* &=& \frac{\epsilon}{8m}\left[\frac{(m+18) (m^2-34m+48)\pm(m-6) M_2}{D_2} \right] \nonumber\\
v^* &=& \frac{\epsilon}{4(m-6)}\left[1+\frac{18}{m}\right]-\frac{3(m-1)}{(m-6)}u^*\nonumber\\
\Delta^* &=& \left[\frac{(m+1)}{2}+\frac{3(m-1)}{(m-6)}\right]u^*-\frac{\epsilon (m-4)}{8(m-6)}\left[1+\frac{18}{m}\right]\nonumber
\end{eqnarray}
where $M_2=( m^4+1740 m^3-48012 m^2+92448 m- 15552)^{1/2}$ and $D_2= (2 m^3-55 m^2 +114m -36)$ 

These pair of fixed points will be called the disordered p-wave fixed points. As can be seen the expressions for the fixed point values are extremely cumbersome. 
So is also the case for the eigenvalues that we obtain as a result of linearizing around this fixed point. 
Thus, for the sake of simplicity, we will discuss the behavior of the system near these fixed points 
for particular representative 
values for the number of order parameter components. 
For small values of $m$, we can show that the fixed point is unphysical. For example,
for $m=3$, it can be shown that the fixed point value is $ u^*=0.291667\epsilon \pm 0.325071\epsilon \, i, v^*=\pm 0.650142\epsilon\, i$ and $\Delta^*=-0.291667\epsilon$, clearly indicating the unphysical nature of the fixed point. 
The fixed points are real only when $m\ge 25.1$. 
Let us look at the behavior of the flow near the disordered p-wave fixed point when we are 
in this regime. For instance, when $m=50$, out of the pair of possible fixed points there is 
only one that is physically relevant. This fixed point is given by $u^*=0.0109089\epsilon, v^*=-0.0287184\epsilon, \Delta^*=0.136896\epsilon$. The  
 eigenvalues obtained by linearizing around this fixed point for the case $m=50$  are given by $-2.49734\epsilon, 0.798418\epsilon \pm 1.57903\epsilon\, i$ and $-\epsilon$, clearly indicating the unstable nature of the 
 fixed point. This state of affairs wherein one gets physically relevant values for the fixed points that 
 are however unstable continues until we reach $m\approx 211$. For $m$ values that are greater 
 than $\approx 211$, we can show that the fixed point value of $\Delta$ is negative, thus implying that disordered $p$-wave fixed point goes once again unphysical. 
  
  In conclusion, in this brief report we have studied the effect of weak impurities on the first order transition found in a p-wave superconductor. 
  It was seen that for the physically relevant 
  situation of small number of order-parameter components (with $m \neq 1$),
   the phase transition is still of the fluctuation driven first order type. 
   However, it was also shown that the case $m=1$, maps itself onto the disordered s-wave 
   superconductor that was studied in Ref. \cite{cardy}. 

   The conclusion that the 
   critical behavior in a random-mass disordered p-wave superconductor remains weakly 
   first order is encapsulated in Fig. \ref{fig:2}, which depicts the numerical solutions of the flow equations, Eq. \ref{eq:1}. These are plotted in Fig. \ref{fig:2}, wherein the case $m=1$, is contrasted with the case $m=2$. For the case $m=2$, we see the runaway flow persists, indicating that the transition remains first order. 
  Fig. \ref{fig:2}, clearly shows that for $m=1$, we flow into the disordered s-wave fixed point of 
  Ref. \cite{cardy}. 
   When the number of 
   order parameter components is large i.e., $m > 211$, the pure fixed point found in \cite{qili} 
   goes stable.    
   It should be added that our results were obtained in a $4-\epsilon$ expansion. However, 
   in the case of the ordinary s-wave superconductors with large values of the Ginzburg parameter, $
   \kappa$,  it has been shown that the $4-\epsilon$ expansion is unable to capture the continuous second order 
   nature of the transition seen by using other techniques, \cite{mc}, \cite{mcb}. Hence, one had to lay recourse to analytical techniques other than the $4-\epsilon$ expansion to capture the continuous 
  transition found in superconductors with large $\kappa$. For instance, in \cite{herbut}, 
 the RG is performed in exactly $d=3$, to capture the continuous phase transition. Efforts are 
  on to perform a similar analysis for the case of the p-wave superconductor \cite{pr}.

\acknowledgments{ We acknowledge valuable discussions with D. Belitz, T. Vojta
C. Janani, S. Govindarajan, P. K. Tripathy, Maria-Teresa Mercaldo, Chitra Nayak and A. Lakshminarayan.

\end{document}